\documentclass{article}

\usepackage[scr=boondoxo,scrscaled=1.05]{mathalfa}
\usepackage[pdftex]{graphicx}
\usepackage{CJKutf8}
\usepackage{amsmath,amssymb,hhline,tikz,mathtools}
\usepackage{enumerate}
\usetikzlibrary{shapes.geometric,decorations.markings}
\usetikzlibrary{calc}
\usepackage{microtype}
\usepackage[title]{appendix}
\usepackage{etoolbox}
\patchcmd{\appendices}{\quad}{: }{}{}
\usepackage{algorithm}
\usepackage{algorithmicx}
\usepackage{algpseudocode}
\usepackage{rotating}
\usepackage{afterpage}
\usepackage{xstring}%%% needed for \rowvec macros
\usepackage{xcolor}
\usepackage{cmap}
\usepackage{mathdots}
\usepackage{colortbl}
\usepackage{booktabs}
\usepackage{textcomp}     % access \textquotesingle
\usetikzlibrary{backgrounds}
\usepackage{pgfplots}
\pgfplotsset{compat=1.12}
\pgfmathsetseed{5112}
\usepackage{multirow}
\usepackage{pgfplotstable}
\usepackage{adjustbox}
\usepackage{diagbox}

\PassOptionsToPackage{hyphens}{url}
\usepackage{hyperref}
\hypersetup{
    pdftitle={Word Embedding Techniques for Malware Classification},
    pdfauthor={Mark Stamp},
    bookmarksnumbered=true,     
    bookmarksopen=true,         
    bookmarksopenlevel=3,       
    colorlinks=true,linkcolor=black,citecolor=black,urlcolor=blue,filecolor=blue,
    pdfstartview=Fit,           
    pdfpagemode=UseOutlines,
    pdfpagelayout=SinglePage
}

\definecolor{darkgreen}{rgb}{0.125,0.5,0.169}

\algrenewcommand{\algorithmiccomment}[1]{{\color{red}{\tt //}\ #1}}
\algnewcommand{\Initialize}[1]{%
  \State \textbf{Initialize:}
  \Statex \hspace*{\algorithmicindent}\parbox[t]{.8\linewidth}{\raggedright #1}
}
\algnewcommand{\Given}[1]{%
  \State \textbf{Given:}
  \Statex \hspace*{\algorithmicindent}\parbox[t]{.8\linewidth}{\raggedright #1}
}

\long\def\symbolfootnotetext[#1]#2{\begingroup%
\def\thefootnote{\fnsymbol{footnote}}\footnotetext[#1]{#2}\endgroup}

\advance\oddsidemargin by -0.15in
\advance\textwidth by 0.3in
\let\oldsqrt\sqrt
\def\sqrt{\mathpalette\DHLhksqrt}
\def\DHLhksqrt#1#2{%
\setbox0=\hbox{$#1\oldsqrt{#2\,}$}\dimen0=\ht0
\advance\dimen0-0.2\ht0
\setbox2=\hbox{\vrule height\ht0 depth -\dimen0}%
{\box0\lower0.4pt\box2}}
\def\clap#1{\hbox to 0pt{\hss#1\hss}}

\allowdisplaybreaks

\def\figureFastSpeed{s}\def\figureSpeed{f}
\let\figureFastSpeed=\figureSpeed
\def\selectFigureSpeed#1#2{
\if\figureSpeed\figureFastSpeed #1\else #2\fi}

\def\srowvecc#1#2{(\!\begin{array}{cc} 
      \noexpandarg\IfBeginWith{#1}{-}{\! #1}{#1}
    & #2\kern-0.5pt\end{array}\!)}
\def\rowvecc#1#2{\left(\!\begin{array}{cc} 
      \noexpandarg\IfBeginWith{#1}{-}{\! #1}{#1}
    & #2\kern-0.5pt\end{array}\!\right)}
\def\rowveccc#1#2#3{\left(\!\begin{array}{ccc} 
      \noexpandarg\IfBeginWith{#1}{-}{\! #1}{#1}
    & #2 
    & #3\kern-0.5pt\end{array}\!\right)}
\def\rowvecccc#1#2#3#4{\left(\!\begin{array}{cccc}
      \noexpandarg\IfBeginWith{#1}{-}{\! #1}{#1}
    & #2 
    & #3 
    & #4\kern-0.5pt\end{array}\!\right)}
\def\srowvecccc#1#2#3#4{\bigl(\!\begin{array}{cccc}
      \noexpandarg\IfBeginWith{#1}{-}{\! #1}{#1}
    & #2 
    & #3 
    & #4\kern-0.5pt\end{array}\!\bigr)}
\def\rowveccccc#1#2#3#4#5{\left(\!\begin{array}{ccccc} 
      \noexpandarg\IfBeginWith{#1}{-}{\! #1}{#1}
    & #2
    & #3
    & #4
    & #5\kern-0.5pt\end{array}\!\right)}
\def\srowvecccccc#1#2#3#4#5#6{(\!\begin{array}{cccccc} 
      \noexpandarg\IfBeginWith{#1}{-}{\! #1}{#1}
    & #2
    & #3
    & #4
    & #5
    & #6\kern-0.5pt\end{array}\!)}
\def\rowvecccccc#1#2#3#4#5#6{\left(\!\begin{array}{cccccc} 
      \noexpandarg\IfBeginWith{#1}{-}{\! #1}{#1}
    & #2
    & #3
    & #4
    & #5
    & #6\kern-0.5pt\end{array}\!\right)}
\def\figureType{*}\def\figureSlowType{slowType}
\def\selectFigureType#1#2{
\if\figureType\figureSlowType #1\else #2\fi}
\makeatletter
\newcommand{\lowsub}[1]{\mathpalette{\raisem@th{#1}}}
\newcommand{\raisem@th}[3]{\raisebox{-#1}{$#2#3$}}
\makeatother
\def\halfthin{\kern 0.083em}

\def\naive{na\"{i}ve}

\def\eref#1{{\color{black}(\ref{#1})}}

\pgfkeys{
    /pgf/number format/precision=2, 
    /pgf/number format/fixed zerofill=true }
    
\pgfplotstableset{
    /color cells/min/.initial=0,
    /color cells/max/.initial=1000,
    /color cells/textcolor/.initial=,
    color cells/.code={%
        \pgfqkeys{/color cells}{#1}%
        \pgfkeysalso{%
            postproc cell content/.code={%
                \begingroup
                \pgfkeysgetvalue{/pgfplots/table/@preprocessed cell content}\value
\ifx\value\empty
\endgroup
\else
                \pgfmathfloatparsenumber{\value}%
                \pgfmathfloattofixed{\pgfmathresult}%
                \let\value=\pgfmathresult
                \pgfplotscolormapaccess
                    [\pgfkeysvalueof{/color cells/min}:\pgfkeysvalueof{/color cells/max}]%
                    {\value}%
                    {\pgfkeysvalueof{/pgfplots/colormap name}}%
                \pgfkeysgetvalue{/pgfplots/table/@cell content}\typesetvalue
                \pgfkeysgetvalue{/color cells/textcolor}\textcolorvalue
                \toks0=\expandafter{\typesetvalue}%
                \xdef\temp{%
                    \noexpand\pgfkeysalso{%
                        @cell content={%
                            \noexpand\cellcolor[rgb]{\pgfmathresult}%
                            \noexpand\definecolor{mapped color}{rgb}{\pgfmathresult}%
                            \ifx\textcolorvalue\empty
                            \else
                                \noexpand\color{\textcolorvalue}%
                            \fi
                            \the\toks0 %
                        }%
                    }%
                }%
                \endgroup
                \temp
\fi
            }%
        }%
    }
}
\makeatletter
\newcommand*\bigcdot{\mathpalette\bigcdot@{.5}}
\newcommand*\bigcdot@[2]{\mathbin{\vcenter{\hbox{\scalebox{#2}{$\m@th#1\bullet$}}}}}
\makeatother

\def\O{{\cal O}}

\def\k{\kern 2.75pt}

\newlength{\xxxxx}
\def\log{\mbox{log}}

\def\z{\phantom{0}}

\newcommand\Tstrut{\rule{0pt}{2.6ex}}         % = `top' strut
\def\sscoin{%
  \leavevmode
  \vtop{\offinterlineskip %\bfseries
    \setbox0=\hbox{\scriptsize S}%
    \setbox2=\hbox to\wd0{\hfil\hskip-.03em
    \vrule height .3ex width .08ex\hskip .08em
    \vrule height .3ex width .08ex\hfil}
    \vbox{\copy2\box0}\box2}}
\newcommand\affil[2]{%
  \begingroup
  \renewcommand\thefootnote{}\footnote{\llap{$\hbox{}^{#1}\hbox{}$}#2}%
  \addtocounter{footnote}{-1}%
  \endgroup
}
\newcommand\markonly[1]{%
$\hbox{}^{\mbox{\kern4.5pt,\kern0.75pt #1}}$
}

\advance\oddsidemargin by -0.525in
\advance\textwidth by 1.05in
\advance\topmargin by -0.625in
\advance\textheight by 1.25in

\title{\vspace{-0.5in}Sentiment Analysis for Troll Detection on Weibo}

\author{
Zidong Jiang\thanks{zidong.jiang@sjsu.edu}\ \ \ \  
Fabio Di Troia\thanks{fabio.ditroia@sjsu.edu}\markonly{\sscoin}\ \ \  
\setcounter{footnote}{3}
Mark Stamp\thanks{mark.stamp@sjsu.edu}\markonly{\sscoin}
}
\date{}

\begin{document}

\maketitle

\vglue-0.35in

\affil{\sscoin}{Department 
of Computer Science,
San Jose State University,
San Jose, California}

\abstract
The impact of social media on the modern world is difficult to overstate. Virtually all 
companies and public figures have social media accounts on popular platforms such 
as Twitter and Facebook. In China, the micro-blogging service provider, 
Sina Weibo, is the most popular such service. To influence public opinion, 
Weibo trolls---the so called Water Army---can be hired to post deceptive comments.
In this paper, we focus on troll detection via sentiment analysis and other user activity 
data on the Sina Weibo platform. %, where the content is mainly in Chinese. 
We implement techniques for Chinese sentence segmentation, word embedding, 
and sentiment score calculation. 
In recent years, troll detection and sentiment analysis have been studied, but we are 
not aware of previous research that considers troll detection based on sentiment analysis. 
We employ the resulting techniques to 
develop and test a sentiment analysis 
approach for troll detection, based on a variety of machine learning strategies. Experimental
results are generated and analyzed. A Chrome extension is presented that implements 
our proposed technique, which enables real-time troll detection when a user browses 
Sina Weibo.

\section{Introduction}

Social media plays a significant role in the ongoing development of the Internet,
as people tend to acquire more information from social media than other platforms. 
Deceptive comments created by trolls present a 
challenging problem in social media applications. Trolls can be hired to publish 
misleading comments in an effort to affect public opinion related to events or people, 
or even to negatively influence the economy of a country. 

Sina Weibo is a widely used micro-blogging social media platform in China. A majority of 
Weibo posts are written in Chinese and, like Twitter, most posts published on Weibo 
are short---until recently, there was a~140 character limit. 
With the number of daily active users in excess of 200 million (as of 2019), 
Weibo is one of the largest social media platforms in China. 

Weibo is based on weak
relationships, in the sense that a user can share content 
that is visible to all of the user base. 
Therefore, many celebrities, businesses, and Internet influencers all over the 
world register as Weibo users in an effort to expand their exposure to the Chinese public. 
Weibo has become a platform where government and businesses can 
communicate more efficiently with the general public.  

The Chinese Water Army refers to a group of people who can be hired to post deceptive 
comments on Weibo. Such troll activity is difficult to detect, due in part to the unsegmented 
characteristic of the Chinese sentences. In some cases, Chinese sentences
can be segmented in different ways to yield different meanings.

Recent research has shown that hidden Markov models (HMM) are effective for 
sentiment analysis of English text~\cite{shopcmt1}. Chinese word segmentation 
can also be accomplished using HMMs~\cite{chseg2,chseg3,chseg1}. In
this research, we use HMMs, Word2Vec and other learning techniques to perform word segmentation 
and sentiment analysis on Sina Weibo ``tweets'' for the purpose of 
detecting potential troll activity. We use Word2Vec and HMMs for Chinese
text segmentation, we employ HMMs and \naive\ Bayes for sentiment analysis,
and we use XGBoost and support vector machines (SVM) for 
troll detection.

%Again, a key point of this research is to apply sentiment analysis to the troll detection problem. 
We have generated a large training dataset by crawling the Sina Weibo and 
Tencent Weibo platforms. Using an HMM-based Chinese sentence segmentation 
model comparable to that in~\cite{chseg1}, we pre-process each post into a list of words. 
Then, following the approach in~\cite{FengFineChinese}, we construct a Word2Vec 
similarity scoring matrix based on the word list that we have generated. 
A baseline of sentiment is determined from the corpus that we have collected. 

For sentiment analysis, we use a Word2Vec based technique to calculate 
sentiment scores.  We use extracted features from Weibo comments 
as observations to train HMM models for each emotion, and we use the trained models 
to determine the emotions of each comment. We use an XGBoost model to aggregate 
sentiment analysis results with user activity data to build the troll detection model.
As a point of comparison, we experiment with an approach based on support vector machines.

Finally, we present a Chrome extension that we have developed. This Chrome extension 
implements our troll detection model, and it
enables us to detect potential troll activity on Weibo in real-time.

The remainder of this paper is organized as follows. In Section~\ref{chap:background}
we discuss relevant background topics. 
Section~\ref{sect:relatedWork} contains an overview of selected
previous work. In Section~\ref{chap:dataset}, we consider 
data sources and data collection methods. 
In Section~\ref{chap:experiment}, we provide implementation details
and includes experimental results. 
Lastly, in Section~\ref{chap:conclusion} we  give a summary of our work, 
including a brief discussion of possible directions for future development. 

\section{Background}\label{chap:background}

In this section, we discuss several relevant background topics.
First, to motivate this research, we discuss trolls in the context
of social media. Then we introduce machine learning models
that are used in this research.
We conclude this section with a brief discussion of the 
evaluation metric that we employ.

\subsection{Trolls}

Troll users publish misleading, offensive or trivial following-up content in online 
communities. The content of a troll posting generally falls into one of several categories. 
It may consist of an apparently foolish contradiction of common knowledge, 
a deliberately offensive insult to the readers of a newsgroup or mailing list, 
or a broad request for trivial follow-up postings. The result of such posting is 
frequently a flood of angry responses. In some cases, the follow-up messages posted 
in response to a troll can constitute a large fraction of the contents of a 
newsgroup or mailing list on a particular topic over an extended period of time. 
These messages may be transmitted around the world to vast numbers of computers, 
wasting network resources and costing resources. 
Troll threads frustrate people who are trying to carry on substantive 
discussions~\cite{trolldefination}.

%One example showing the activity of trolls occurred during 
%the 2016 U.S. presidential election. Suspected troll users on Twitter 
%and Reddit posted repetitive and negative comments related to the 
%Democratic party candidate. Most of those accounts were associated with 
%Russian government–sponsored propaganda operations~\cite{russiantroll2016}. 

Organized troll activity on the Sina Weibo platform was first detected in~2013. 
This initial group of troll users consisted of about~20,000 individuals 
in~50 ICQ chat groups associated with a person nicknamed ``Daxia.'' 
Subsequently, troll activity became an online business on the Weibo platform. 
Trolls can be hired by businesses to publish negative comments against their 
competitors or to generate anonymous good reviews or positive comments. 
Prior to~2015, much of the troll activity on Weibo was designed to adversely 
affect the reputation of businesses.
After 2015, stricter controls were set on speech on the Internet in China,
and Sina Weibo developed a more sophisticated infrastructure to filter such troll comments. 
Currently, most of the troll activity on Weibo turned is designed
to promote celebrities and companies. 

\begin{figure}[!htb]
    \centering
    \includegraphics[width=0.5\textwidth]{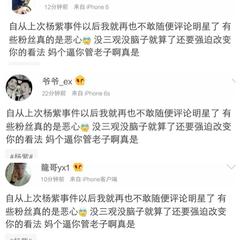}
    \caption{Weibo comments dominated by troll activity~\cite{yangzitrolldomination}}
    \label{fig:shuaping}
\end{figure}

Troll users on the Weibo platform can be categorized by their source of content. 
Traditionally, trolls use automated fake accounts to post repeated messages 
in an effort to dominate the comments. An example of such activity is shown 
in Figure~\ref{fig:shuaping}. However, the Weibo 
platform has recently improved their infrastructure to block these repeated messages 
from users, based on proxy detection, combined with message filters for repeated comments. 

Recently, troll users have become more sophisticated. Some organized Weibo trolls 
are supervised by a management 
group who controls what, when, and where they reply on the Weibo platform.  
Specific details of the comments that each troll account publishes are made by individual 
troll users rather than being copied from the management group. The management 
group only gives out the overall emotional trend that the comments should convey. 
Thus, content made by troll users are repetitive but not monotonously so. 
This fact makes troll detection on Weibo challenging, since troll comments 
are composed and published by real human users. Furthermore, in recent years, 
trolls are mostly hired by companies and celebrities to make positive 
comments towards themselves. 

%\subsection{Machine Learning for Troll Detection}

\subsection{Machine Learning Techniques}

Content-based troll detection usually utilizes natural language processing (NLP)
using machine learning to analyze and categorize text.
This is accomplished by constructing language processing models on 
comments and posts so as to label comments with a high polarity of emotion
or repetitiveness. By applying sentiment analysis methods, we can filter 
comments with either high or low sentiment scores representing extreme positive 
or negative sentiment. This is accomplished by calculating word relevance, 
and by analyzing correlations using word embedding techniques, 
such as Word2Vec. We can then mark potential troll comments or pass along
user information behind such comments to the next stage of a troll detection model. 
A key point of this research is to use sentiment analysis in
troll detection. 

Classifying specific comments as troll activity is challenging. Therefore,
utilizing user behavioral information to discern deceptive activity 
is a popular trend in troll detection. Like most social media platforms, 
Weibo has numerous user relationship data, such as the number of followers, 
number that a user is following, user rank, and number of original Weibo tweets. 
Also, trolls commonly make attacks in a small time window following a 
specific tweet~\cite{weibospammer}. We can utilize this fact in combination with 
other user relationship information in a troll detection system. The goal 
of including such data in our troll detection approach is to reduce 
the false negatives that affect strictly content-based detection methods.

Next, we introduce the various machine learning techniques used in this
research. Specifically, we discuss hidden Markov models (HMM), Word2Vec,
XGBoost, and support vector machines (SVM) in some detail.

\subsubsection{Hidden Markov Models}

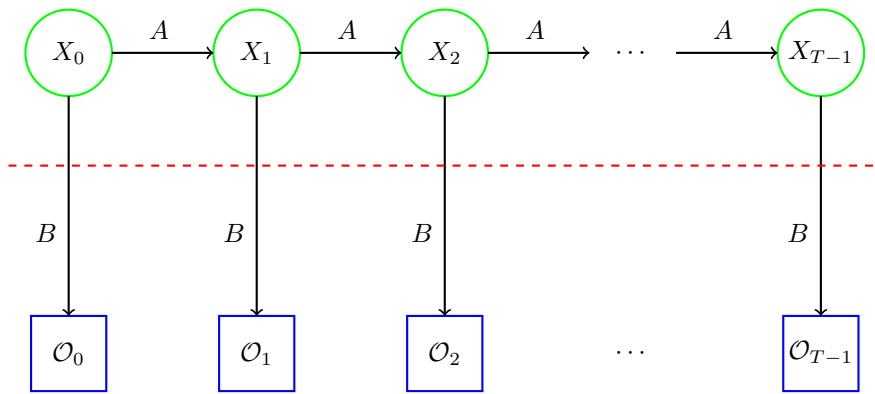
\begin{figure}[!htb]
\centering
  \begin{tikzpicture}[scale=1.0]
    
    % squares
    \draw[thick,color=blue] (0,0) rectangle (1,1);
    \draw[thick,color=blue] (2.5,0) rectangle (3.5,1);
    \draw[thick,color=blue] (5,0) rectangle (6,1);
    \draw[thick,color=blue] (10,0) rectangle (11,1);

    % circles
    \draw[thick,color=green] (0.5,4.5) circle (0.575);
    \draw[thick,color=green] (3,4.5) circle (0.575);
    \draw[thick,color=green] (5.5,4.5) circle (0.575);
    \draw[thick,color=green] (10.5,4.5) circle (0.575);
    
    % observations
%    \node at (-1.5,0.5){Observations:};
    \node at (0.5,0.5){$\O_0$};
    \node at (3,0.5){$\O_1$};
    \node at (5.5,0.5){$\O_2$};
    \node at (8,0.5){$\cdots$};
    \node at (10.5,0.5){$\O_{T-1}$};

    % States
%    \node at (-1.75,4.5){Markov process:};
%    \node at (0.5,4.5){$X_0$};
%    \node at (3,4.5){$X_1$};
%    \node at (5.5,4.5){$X_2$};
%    \node at (8,4.5){$\cdots$};
%    \node at (10.5,4.5){$X_{T-1}$};
    \node at (0.5,4.5){$X_0$};
    \node at (3,4.5){$X_1$};
    \node at (5.5,4.5){$X_2$};
    \node at (8,4.5){$\cdots$};
    \node at (10.5,4.5){$X_{T-1}$};
       
    % A's
    \node at (1.7,4.8){$A$};
    \node at (4.2,4.8){$A$};
    \node at (6.7,4.8){$A$};
    \node at (9.2,4.8){$A$};
    
    % B's
    \node at (0.2,2.1){$B$};
    \node at (2.7,2.1){$B$};
    \node at (5.2,2.1){$B$};
    \node at (10.2,2.1){$B$};
    
    % circle to circle
     \draw[thick,color=black,->] (1.075,4.5) -- (2.425,4.5);
     \draw[thick,color=black,->] (3.575,4.5) -- (4.925,4.5);
     \draw[thick,color=black,->] (6.075,4.5) -- (7.425,4.5);
     \draw[thick,color=black,->] (8.575,4.5) -- (9.925,4.5);

    % circle to square
     \draw[thick,color=black,->] (0.5,3.925) -- (0.5,1);
     \draw[thick,color=black,->] (3.0,3.925) -- (3.0,1);
     \draw[thick,color=black,->] (5.5,3.925) -- (5.5,1);
     \draw[thick,color=black,->] (10.5,3.925) -- (10.5,1);

    % curtain
    \draw[thick,dashed,color=red] (-0.3,3) -- (11.2,3);
   
\end{tikzpicture}
\caption{Hidden Markov model}\label{fig:hmm}
\end{figure}

Hidden Markov models (HMMs) are well-known for their use in 
pattern prediction and for deriving hidden states from observations. 
An HMM (of order one) is a stochastic model representing states where each future state 
depends only on the current state, and not on states further in the past. By training 
an HMM on an observation sequence, we can obtain the probability of the transitions 
between each hidden state and probability distributions for the observations, based
on those hidden states. A generic HMM is illustrated in Figure~\ref{fig:hmm}. 

In Figure~\ref{fig:hmm}, the matrix~$A$ drives the Markov process
for the hidden states, while the matrix~$B$ probabilistically relates
the hidden states to the observations~$\O_i$.
The HMM notation is summarized in Table~\ref{tab:HMMnotation}. 

\begin{table}[!htb]
\caption{HMM notation}\label{tab:HMMnotation}
\vglue 0.1in
\centering
  \begin{tabular}{cl} \midrule\midrule
    \textbf{Notation} & \hspace*{0.5in}\textbf{Description}\\ \midrule
    $T$ & Length of the observation sequence\\
    $N$ & Number of states in the model\\
    $M$ & Number of observation symbols\\
    $Q$ & Distinct states of the Markov process, $q_0,q_1,\ldots,q_{N-1}$\\
    $V$ & Possible observations, assumed to be $0,1,\ldots,M-1$\\
    $A$ & State transition probabilities\\
    $B$ & Observation probability matrix\\
    $\pi$ & Initial state distribution\\
    $\O$ & Observation sequence, $\O_0,\O_1,\ldots,\O_{T-1}$ \\ \midrule\midrule
  \end{tabular}
\end{table}

Applications of HMMs are extremely diverse, but for our purposes, two
relevant uses are English text analysis and 
speech recognition~\cite{revealhmm}. Other applications of 
HMMs range from classic cryptanalysis~\cite{hmmcipher} to malware
detection~\cite{Dhanasekar2018,hmmmetamalware}. 
In this research, we use HMMs for both Chinese word
segmentation and for emotion classification.

%Sentiment analysis is a classic topic in natural language processing (NLP)
%where machine learning is used to find the subjective attitude of authors 
%or speakers. In this paper, we use this technique to categorize 
%Weibo tweet sentiments as anger, disgust, fear, joy, sadness, or surprise~\cite{emotionCate}. 

%Previous research suggests that the best models for converting segmented 
%text into a categorical sentiment rely on word embedding techniques
%as a means of calculating sentiment scores. 
%In Figure~\ref{fig:chineseSentiAnalysisProc}, 
%we outline the process of Chinese sentiment analysis 
%that we implement in this research. 
%
%\begin{figure}[!htb]
%    \centering
%    \includegraphics[width=0.25\textwidth]{images/workflow.png}
%    \caption{Sentiment analysis procedure for Chinese language}
%    \label{fig:chineseSentiAnalysisProc}
%\end{figure}

\subsubsection{Word2Vec}

Word2Vec has recently gained considerable popularity in 
natural language processing (NLP)~\cite{word2vecref}.
This word embedding technique is based on a shallow neural network, with the weights
of the trained model serving as embedding vectors---the trained model itself serves no
other purpose. These embedding vectors capture significant relationships 
between words in the training set. Word2Vec can also
be used beyond the NLP context to model relationships between 
more general features or observations.

When training a Word2Vec model, we must specify the desired
vector length, which we denote as~$N$. Another key parameter
is the window length~$W$, which represents the width of a sliding
window that is used to extract training samples from the data.

Certain algebraic properties hold for Word2Vec
embeddings. For example, suppose that we train a state-of-the-art
Word2Vec model on English text. Further, suppose that we let
$$
  w_0=\mbox{``king''}, w_1=\mbox{``man''}, w_2=\mbox{``woman''}, w_3=\mbox{``queen''} ,
$$
and we define~$V(w_i)$ to be the Word2Vec embedding of word~$w_i$.
Then according to~\cite{word2vecref}, the vector~$V(w_3)$
is closest to
$$
  V(w_0) - V(w_1) + V(w_2)
$$
where ``closeness'' is in terms of cosine similarity.
Results such as this indicate that in the NLP context,
Word2Vec embeddings capture meaningful
aspects of the semantics of the language.

In this research, we train Word2Vec models Chinese text.
These models are then used for sentiment analysis
of Weibo tweets.

\subsubsection{XGBoost}

Boosting is a general technique for constructing
a stronger classifier from a large collection
of relatively weak classifiers~\cite{ChenG16}. 
XGBoost typically uses decision trees as the base classifiers. 
To generate our models, we use the XGBoost package in Python. With this
implementation, it is easy to analyze the significance of each individual feature 
relative to the overall model, and thus eliminate ineffective features.

\subsubsection{SVM}

In~\cite{Bennett2000SupportVM}, 
support vector machine (SVM) is described 
as ``a rare example of a methodology where geometric intuition, 
elegant mathematics, theoretical guarantees, and practical algorithms meet.''
The essential ideas behind SVMs are the following.
\begin{description}
\item[Separating hyperplane]\hspace*{-9pt}--- We seek a hyperplane that will separate two 
labeled classes.
\item[Maximize the margin]\hspace*{-9pt}--- We want to find hyperplane 
that maximizes the ``margin'' between two classes, where 
margin is defined as the minimum distance. 
\item[Work in higher dimensional space]\hspace*{-9pt}--- By shifting the problem to a higher dimensional
space, there is a better chance that we can find a separation hyperplane, or
separating hyperplane with a larger margin.
\item[Kernel trick]\hspace*{-9pt}--- Perhaps surprisingly, we are able to work in a higher dimensional space
without paying any significant penalty with respect to computational complexity.
This is a powerful ``trick'' and is the key reason why SVM is one of the most 
popular machine learning techniques available.
\end{description}

In this research, SVM serves as a comparison to XGBoost for troll detection. 
We find that XGBoost performs better on one of our datasets, while SVM
is superior on another dataset. 

\subsection{Evaluation Metric}

We use accuracy as the primary measure of success for all of our classification experiments. 
The accuracy is computed as
$$
\mbox{accuracy} = \frac{\mbox{TP}+\mbox{TN}}{\mbox{TP}+\mbox{TN}+\mbox{FP}+\mbox{FN}} ,
$$
where TP is number of true positive cases, TN is true negatives, FP is false positives, 
and FN is the number of false negatives. Accuracy can be seen to simply be
the ratio of correct classifications to the total number of classifications.

%We also consider receiver operating characteristic (ROC) curves 
%when analyzing our experimental results~\cite{roccurve}. An ROC 
%curve is constructed by plotting the true positive rate (TPR) versus
%the false positive rate (FPR) as the threshold passes thru the range of 
%possible values. The area under the ROC curve (AUC), ranging between~0 
%and~1, with the AUC giving the probability that a randomly selected
%positive instance scores higher than a randomly selected negative instance.
%An AUC of~0.5 implies that the binary classifier is not better than 
%flipping a coin, while and AUC that is less than~0.5 indicates that
%the classification criteria should be flipped. One advantage of the AUC
%as compared to accuracy is that the ROC curve does not depend on
%a specific threshold, since the AUC accounts for all possible thresholds.

Finally, we note that cross validation is used in all of our experiments.
Cross validation is a popular technique that serves to smooth any bias in
the data, while also maximizing the number of datapoints.
Specifically, we employ 5-fold cross validation.

\section{Related Work}\label{sect:relatedWork}

Related work in sentiment analysis includes~\cite{shopcmt2}, 
where a combination of emotional orientation and logistical regression is used to 
analyze Amazon.com reviews. By filtering the training dataset by text length, 
vocabulary complexity, correlation with the product, sentiment similarity, and transition words, 
the proposed model achieved~91.2\%\ accuracy. 
For the problem of fake Weibo tweet detection---as opposed to the troll detection
we consider in this research---an XGBoost model based 
on user activity achieved~93\% accuracy in~\cite{liuwanglong}. 

From our review of the literature, it appears that only~\cite{seah} 
applies sentiment analysis to the troll detection problem. %As mentioned above,
The work in~\cite{seah} applies domain adaptation techniques to a 
recursive neural tensor network (RNTN) sentiment analysis model to detect trolls
that post repetitive, destructive, or deceptive comments.
This previous work achieves~78\%\ accuracy. 
The results in~\cite{seah} serve as a baseline for our research. 

Sentiment analysis is widely used for mining subjective information in 
online posts. In~\cite{hmmsyntactic}, Kim, et al., use hidden Markov models 
with syntactic and sentiment information for sentiment analysis of Twitter data. 
This differs from classic approaches that use $n$-grams and polarity lexicons, 
as they group words based on similar syntactic and sentiment groups (SIG), 
then build HMMs, where the SIGs define the hidden states. Zhao and Ohsawa~\cite{shopcmt1}  
propose a two-dimensional HMM to analyze Amazon reviews in Japanese.
For our purposes, this work illustrates an important method for converting 
segmented Japanese text into word vectors using Word2Vec. 
Feng and Durdyev~\cite{FengFineChinese} implemented three types of 
classification models (SVM, XGBoost, LSTM) for the aspect-level sentiment analysis 
of restaurant customer reviews in Chinese. 
According to the research in~~\cite{FengFineChinese}, 
LSTM yields better F-1 scores and accuracy, as compared to SVM and XGBoost. 
Further related research can be found in Liu, et al.\cite{LiuLuo2015}, 
which uses a self-adaptive HMM. 

%As far as the authors are aware, the only research applying sentiment analysis
%to troll detection is that of Seah, et al~\cite{seah}. They apply domain adaptation 
%techniques to recursive neural tensor network (RNTN) sentiment analysis model 
%to detect repetitive, destructive, and deceptive forum posts, 
%mainly written in colloquial Singaporean English. In this work, 
%classification accuracy of~78\%\ accuracy is attained. 

Troll detection based on user characteristic data, is considered in 
Zhang, et al.~\cite{weibospammer}. In this paper, Weibo troll detection is 
based on a Bayesian model and genetic algorithm. The proposed technique 
includes novel features (as compared to previous work) such as the ratio of followers, 
average posts, and Weibo credibility, and achieves an accuracy of about~90\%.

Liu, Wang and Long~\cite{liuwanglong} use XGBoost to detect fake 
Weibo posts based on features such as a user's number of posts, description, 
gender, followers, and reposts. The authors attain an accuracy of more than~95\%. 
Both~\cite{weibospammer} and~\cite{liuwanglong} use data beyond Weibo post 
text itself, and achieve good results. This previous work serves as inspiration for some of
the features considered in this paper.

The special interest group for Chinese language processing (SIGHAN) 
of the Association for Computational Linguistics
organizes competitions for Chinese word segmentation. 
In the first SIGHAN bake-off event in~2003, Zhang, et al.~\cite{zhang-etal-2003-hhmm} 
proposed a word segmentation approach using hierarchical HMMs 
to form a Chinese lexical analyzer, ICTCLAS. In 2005, 
Masayuki, et al.~\cite{asahara-etal-2005-combination} presented three word segmentation 
models, including a character tagging classifier based on 
support vector machines (SVM)  that also used maximum expropriation Markov models 
and conditional random fields. These models were based on previously proposed 
methods, with a different combination of out-of-vocabulary (OOV) extraction 
techniques being used. 

In general, for Chinese word segmentation,
character-based models perform better than word-based models. 
Wang, Zong, and Su~\cite{wang-etal-2009-suitable} highlighted that 
OOV techniques for word extraction performs poor 
for in-vocabulary (IV) words. They proposed a generative model that 
performs well on both OOV and IV words, and achieved good results 
on the popular SIGHAN datasets. Chen, Chang, and Pei~\cite{chen-etal-2014-joint} 
report the use of Gibbs sampling in combination with both word-based hierarchical 
process models and character-based HMMs. Their solution achieved better 
performance (in terms of~$F_1$ score) than the state-of-the-art models at that time.

\section{Datasets}\label{chap:dataset}

We acquired data and generated additional data for the various parts 
of this research. We have Chinese segmentation data, 
sentiment analysis data, data consisting of Weibo comments, 
and user data corresponding to the Weibo comments data. 
This data is split into three datasets, namely,
a Chinese segmentation dataset,
a sentiment analysis dataset,
and a troll detection dataset.
Next, we discuss each of these three datasets.

\subsection{Chinese Segmentation Dataset}\label{sect:chinesesegdata}

For Chinese sentence segmentation, we acquired the dataset used in
the SIGHAN 2005 Competition for Chinese sentence processing~\cite{sighan2}. 
This dataset includes training, testing, and validation data. The training data 
consists of approximately~860,000 segmented Chinese sentences. Most of these 
sentences are from newspapers and published books. The test set includes 
about~22,000 unsegmented sentences from similar sources, while the validation 
set contains the segmentation of all of the sentences in the test set. 
Table~\ref{tab:chinesesegcorp} gives additional statistics for this dataset. 

\begin{table}[!htb]
    \caption{Chinese segmentation dataset (SIGHAN 2nd Bakeoff 2005)}
    \label{tab:chinesesegcorp}
    \vglue 0.1in
    \centering
    \begin{tabular}{c|c|c} \midrule\midrule
        \textbf{Source} & \textbf{Training} & \textbf{Testing} \\ \midrule
        Academia Sincia & 708,953 & 14,432 \\
        Peking University & \z19,056 & \z1,944 \\
        City University of Hong Kong & \z53,019 & \z1,492 \\
        Microsoft Research Asia & \z86,924 & \z3,985 \\
    \midrule\midrule
    \end{tabular}
\end{table}

%\subsubsection{Features and Extraction}\label{sect:FE}

We consider the character-based generative model proposed in~\cite{chseg1}.
In this model, the features from the training data consist of the positions of each character 
in each segmented word. The beginning character in each segment is 
marked as~$B$, any middle character or characters are marked as~$M$, and the ending 
character is marked as~$E$. On the other hand, all one-character words are marked as~$S$. 

\begin{figure}[!htb]
    \centering
    \includegraphics[width=0.5\textwidth]{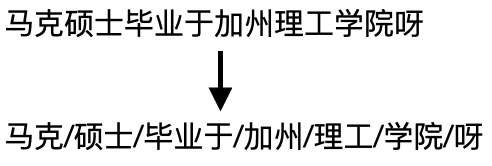}
    \caption{Sample Chinese sentence segmentation}
    \label{fig:chseg}
\end{figure}

For example, consider the sample Chinese sentence in Figure~\ref{fig:chseg},
which includes the correct segmentation for this sentence.
Table~\ref{tab:chsegtable} gives the states corresponding to the sentence 
in Figure~\ref{fig:chseg}.

\begin{CJK*}{UTF8}{gbsn}
\begin{table}[!htb]
    \caption{State sequence for Chinese sentence}\label{tab:chsegtable}
    \vglue 0.1in
    \centering
    \begin{tabular}{c|c|c|c|c|c|c|c|c|c|c|c|c|c} \midrule\midrule
        马 & 克 & 硕 & 士 & 毕 & 业 & 于 & 加 & 州 & 理 & 工 & 学 & 院 & 呀 \Tstrut\\ \midrule
        B & E & B & E & B & M & E & B & E & B & E & B & E & S \\
        \midrule\midrule
    \end{tabular}
\end{table}
\end{CJK*}

\subsection{Sentiment Analysis Dataset}\label{sect:sentidata}

For sentiment analysis, we use the sentiment training dataset from 
the Python SnowNLP package~\cite{snownlp}. This particular dataset 
includes~16,548 
sentences with positive sentiment and~18,574 with negative sentiment. 
The source for this dataset is Chinese online shopping, movie, and book reviews. 
However, this data might not accurately represents tweets and comments 
appearing in Weibo. 

Since there is no public datasets for Weibo,
we crawled about~5 million Sina Weibo posts to obtain additional data 
for our sentiment model. This data includes terms and slang that are
%%%%% How was this created ????? How accurate is it ?????
commonly seen on Weibo. From this data, we created a collection of~2,325,644 
sentences with positive sentiment and~960,899 sentences with negative sentiment. 

From all of our Weibo crawled data, we manually extracted~500 tweets for each of 
the six emotions of interest, namely, happiness, surprise, fear, anger, disgust, 
and sadness~\cite{emotionCate}. This data will be used to train
an HMM for each emotion. 

We process each comment using the Pandas package in Python to remove stop words, 
numbers, nonsense emoji, and single-word comments. A language detection method 
was implemented to detect non-Chinese comments and translate English comments 
into Chinese using the Translator package in Google Translate. We eliminate all 
comments that are in languages other than Chinese and English, which results in
a negligible loss of data. In addition, we removed pure re-posts and tagging 
that is not relevant to our analysis.

%\subsubsection{Features and Extraction}\label{sect:FEsentidata}

For positive and negative sentiment analysis, we used the Word2Vec embedding 
method. The resulting embedding vectors enable us to calculate a word 
sentiment score, after segmenting a Weibo comment into a list of word. 
For sentiment analysis (based on six basic emotions), we use 
the features introduced in~\cite{LiuLuo2015}, which are then used to 
train HMMs for sentiment classification. These features are
mutual information,  $\chi^2$ distance, and
term frequency inverse document frequency,
which are defined as follows.

\begin{description}
\item[MI]\hspace*{-9pt}--- Mutual information (MI) is based on 
correlations between two terms. In this research, MI is used to determine the 
relevance between words and emotions. The formula for MI representing 
the correlation between emotion~$e$ and text~$t$ is given by
$$
    \mbox{MI}(t, e)=\log\frac{P(t\,|\,e)}{P(e)} .
$$
    
\item[CHI]\hspace*{-9pt}--- We use a~$\chi^2$ distance measure (CHI) to quantify
the dependence between emotion~$e$ and text~$t$. The higher the~CHI value, 
the more dependent the text~$t$ is on the emotion~$e$. We calculate CHI as 
%\begin{equation}\label{eq:CHI}
$$
    \mbox{CHI}(t, e)=\frac{N(AD - BC)^{2}}{(A + B)(C + D)(A + C)(B + D)} ,
$$
%\end{equation}
where~$A$ is the prevalence of word~$t$ in comments with emotion~$e$, 
$B$ is the prevalence of word~$t$ in comments with emotions other than~$e$, 
$C$ is the absence of word~$t$ in comments with emotion~$e$,
$D$ is the absence of word~$t$ in comments with emotions 
other than~$e$, and~$N$ is the total number of comments.
    
\item[TF-IDF]\hspace*{-9pt}---  Term frequency inverse document frequency (TF-IDF) was originally 
developed to extract key words from text, for purposes such as indexing. 
We use TF-IDF to determine key words
with respect to the various emotions under consideration.
We compute the TF-IDF as 
$$
    \mbox{TF-IDF}(t, e) = \frac{N_{e,t}}{\displaystyle\sum_{k} N_{k,t}} \log{\left(\frac{N}{n_{e}} + 0.01\right)} ,
$$
where $N_{e,t}$ is the number of times word~$t$ appears in a comment with 
emotion~$e$, $N$ is the total number of comments, and~$n_{e}$ is the
number of comments in which the emotion~$e$ appears.

\end{description}

\subsection{Troll Detection Dataset}\label{sect:trolldata}

After some initial experiments, we realized that there are limitations 
to the features specified in~\cite{seah} and~\cite{LiuLuo2015}. Therefore, 
we introduced more user information related features that we obtained by 
mining Weibo comment data. When crawling the Weibo data, we use the 
JavaScript object notation (JSON) packet 
returned from representational state transfer (REST) calls to the 
Weibo mobile site~\cite{weibomobile}, 
which includes user-related information. 
Typical operations under the REST API include GET, POST, UPDATE, 
and DELETE. We extract the user 
information listed in Table~\ref{tab:weibofield}. We also include a small dataset 
of~673 normal users and~75 trolls from a Kaggle data source~\cite{fakeweiboaccount}.

\begin{table}[!htb]
    \caption{List of user related information from comment data}\label{tab:weibofield}
    \vglue 0.1in
    \centering
%    \resizebox{\textwidth}{!}{
    \begin{tabular}{c|c|c} \midrule\midrule
        \textbf{Field Name} & \textbf{Dataset} & \textbf{Description} \\ \midrule
        \texttt{uid} & UID & Unique User ID for User Account in Weibo  \\
        \texttt{screen\_name} & Username & Displayed User Nickname \\
        \texttt{followers\_count} & Follower & User's follower count \\
        \texttt{follow\_count} & Following & User's following count \\
        \texttt{status\_count} & Original\_post & User's original composed tweet count \\
        \texttt{urank} & User\_rank & User's rated rank by user activity in Weibo \\
        \texttt{verified} & Verified & Whether user is verified celebrity or business \\
        \texttt{description} & Description & User's own description in headline \\
        \texttt{like\_count} & Like\_count & Like count of this comment \\
        \texttt{floor\_number} & Floor\_number & Location where the comment is at \\
        \texttt{text} & Comment & Comment content \\
    \midrule\midrule
    \end{tabular}
%    }
\end{table}

All user information and corresponding comments are grouped by original 
tweet~ID and stored in CSV format. One CSV file contains all of the 
comments regarding one tweet, and each entry represents all of the information 
listed in Table~\ref{tab:weibofield}. We selected eight tweets with a total of~31,980 
comments from Sina Weibo accounts belonging primarily to business owners
and celebrities. The detailed tweet information and statistics 
for this dataset are listed in Table~\ref{tab:commentsdata}.

\begin{table}[!htb]
    \caption{Statistics of troll detection tweets and comments crawled from Weibo}
    \label{tab:commentsdata}
    \vglue 0.1in
    \centering
    \resizebox{0.85\textwidth}{!}{
    \begin{tabular}{c|c|c|r} \midrule\midrule
       \textbf{} & \textbf{Tweet ID} & \textbf{User details} & \textbf{Number} \\ \midrule
        1 & 44275283 & LeEco CEO YT Jia declared bankrupt & 812\z \\
        2 & 44317480 & Actress Yiyan Jiang volunteered teaching in rural & 829\z \\
        3 & 44564209 & Yong actress Zi Yang suspected done plastic surgery & 335\z \\
        4 & 44718878 & Reporting fraud in singer Hong Han Foundation receiving donation 
        		& 1210\z \\
        5 & 44651702 & Singer Hong Han Foundation donation to Wu Han Coronavirus battle 
        		& 3379\z \\
        6 & 44650056 & Criticism of multiple celebrities' donation to Coronavirus battle & 814\z \\
        7 & 43961306 & Suspected breakup of Han Lu and Xiaotong Guan (Actor/ress) & 8371\z \\
        8 & 43961306 & Han Lu and Xiaotong Guan (Actor/ress) showoff their same sweatshirt 
        		& 16,230\z \\
    \midrule\midrule
    \end{tabular}
    }
\end{table}

We manually labeled the data for rows~$1$, $2$, $3$, and~$6$ 
in Table~\ref{tab:commentsdata} as troll or non-troll
by examining the content of each comment.
%Troll users are specified as~1 while non-troll users are labeled~0. 
Combining these results with fake account data from~\cite{fakeweiboaccount}, 
we have about~3500 comment entries for our initial training and testing data 
for the troll detection model. This manual labeling is extremely tedious. 
To accelerate this process, we created a bot based on Selenium~\cite{selenium} 
to help open each user's Weibo page based on the UID that we provide from the dataset.

%\subsubsection{Feature Extraction}

To extract features from the data listed in Table~\ref{tab:weibofield},
we use Python Pandas~\cite{pandas} 
and Numpy~\cite{numpy}. Some of the available features proved to be 
of little use, and these were dropped, as discussed below. 
In order to have better features for our models,
we also perform some feature engineering. For example, 
we note whether users provide a self-description or not. 

Features such as follower count, following count, and the number of
original composed tweets clearly have a high significance in our analysis. 
However, we found that building models with quantitative numbers from these categories
biases the model, due to the large differences across users. 
Weibo users typically only follow a fairly small number of accounts,
while troll users typically follow a large number of accounts. 
Therefore, we dropped the follower and following count in the raw dataset and 
instead compute the ratio of following to follower and use this as a feature. Similarly,
we introduce a feature consisting of the ratio of original posts to followers to help identify
troll users, who often make a large number of posts without a 
commensurate increase in their follower count---we use this engineered feature
in place of the composed post feature from the raw data. 

When crawling Weibo, we noticed that some users frequently comment on the same tweet 
rather than replying to other comments under a tweet. Therefore, we select users who have
more than one comment under a tweet and we count the comments made
for each such user. Then we computed the median of these comment counts. 
Following this approach, a ``frequent comment'' feature is generated based
on users who made more comments than the median number. 

%\subsubsection{Features}

%We use Python and Pandas to extract the data specified in 
%Table~\ref{tab:weibofield}. Initially, we extract as many features as possible from 
%the raw data. We then perform various data manipulations with Pandas and NumPy 
%to engineer more informative features, as discussed above. 

We have a total of~19 features that we use in our XGBoost model. 
One of the engineered features related to the sentiment score is denoted as
\texttt{diffOriginalSenti}. This feature is the score for a comment minus
the sentiment score of the original tweet. 
Table~\ref{tab:alltrollfeature} lists the complete set of features that we obtain
by combining sentiment analysis result and user information data in Table~\ref{tab:weibofield}.

\begin{table}[!htb]
    \caption{Features considered for troll detection model}\label{tab:alltrollfeature}
    \vglue 0.1in
    \centering
    \resizebox{0.85\textwidth}{!}{
    \begin{tabular}{r|c|c|c} \midrule\midrule
           & \textbf{Feature} & \textbf{Description} & \textbf{Source}  \\ \midrule
        F0 & \texttt{follower} & Follower count & Crawled Weibo dataset \\
        F1 & \texttt{following} & Following count & Crawled Weibo dataset \\
        F2 & \texttt{original\_post} & Number of original tweets & Crawled Weibo dataset \\
        F3 & \texttt{urank} & Rank by user activity in Weibo & Crawled Weibo dataset \\
        F4 & \texttt{verified} & User certified or not & Crawled Weibo dataset \\
        F5 & \texttt{like\_count} & Like count for a comment & Crawled Weibo dataset \\
        F6 & \texttt{floor\_number} & Comment location & Crawled Weibo dataset \\
        F7 & \texttt{description} & Self description (1 or 0) & Engineered feature \\
        F8 & \texttt{freqComment} & Frequent comments & Engineered feature \\
        F9 & \texttt{ffRatio} &   \texttt{following} divided by \texttt{follower} & Engineered feature \\
        F10 & \texttt{foRatio} & \texttt{original\_post} divided by \texttt{follower} 
              & Engineered feature \\
        F11 & \texttt{sentiment} & Comment sentiment score (0 to 1) & Engineered feature \\
        F12 & \texttt{diffOriginalSenti} & \texttt{sentiment} minus sentiment of original 
        			& Engineered feature \\
        F13 & \texttt{happy} & Happiness score (0 to 1) & Engineered feature \\
        F14 & \texttt{sad} & Sadness score (0 to 1) & Engineered feature \\
        F15 & \texttt{anger} & Anger score (0 to 1) & Engineered feature \\
        F16 & \texttt{disgust} & Disgust score (0 to 1) & Engineered feature \\
        F17 & \texttt{fear} & Fear score (0 to 1) & Engineered feature \\
        F18 & \texttt{surprise} & Surprise score (0 to 1) & Engineered feature \\
    \midrule\midrule
    \end{tabular}
    }
\end{table}

We would like to 
maximize our troll detection accuracy while minimizing the number of
features needed. To achieve this, we perform feature analysis in order to 
rank the significance of features, so that we can drop features. 
This feature reduction process is discussed below
in Section~\ref{sect:XGB_SVM}.

\section{Implementation and Results}\label{chap:experiment}

In this section, we give our results. First, we discuss the
Weibo crawler that we have implemented. Then we consider our Chinese
word segmentation results, followed by the our emotion
classification technique, both of which are based on hidden Markov models.
Then we consider our Word2Vec based sentiment score.
and our XGBoost and SVM based troll detection results.
We conclude this section with a discussion of a Chrome
extension that implements this troll detection system.

\subsection{Weibo Crawler}

As mentioned above, in order to have sufficient 
training and testing data, we developed a crawler 
to obtain such data directly from the Weibo platform. 
Our crawler extracts posts, comments, and user information. 

To extract posts, the crawler certain considers a number of tweets under 
specific Weibo accounts. Note that Weibo tweets are similar to Twitter tweets, 
in that users can retweet others users' posts to their own Weibo account. 
The crawler disregards retweets and only keeps original posts. 

Comment crawling is used to obtain additional information related to posts.
Most comments contain repetitive messages and include username and hash-tags. 
We remove this extraneous data with the Python Panda Dataframe function 
before pipelining the comments into the word segmentation stage. Also, it is very 
common to see bilingual comments in Weibo, where most of the text
is Chinese, but some English is included. Therefore, we incorporate 
a language detection module extended from the Google language library,
which uses \naive\ Bayes to filter and translate English to Chinese.

Our comment crawler works on Weibo mobile~\cite{weibomobile} data,
where the tweets and comments page are slightly simplified. 
The crawler makes HTTP requests such as 
$$
\mbox{\footnotesize\texttt{https://m.weibo.cn/comments/hotflow?id=TWEETID\&mid=TWEETID\&max\_id=}} 
$$
which yields JSON data from the Weibo platform containing comments related to 
the specified tweet.\footnote{One obstacle we encountered 
was a change in the Weibo mobile site at the 
beginning of~2020. To avoid being blocked when crawling a large number of 
comments, we were forced to modify the crawler to use 
the ``max\_ID'' property for the current comment page.}
We parse the resulting comment data contained in the JSON packet 
to extract all of the raw data, including user information and comment content
using the BeautifulSoup package in Python~\cite{BS}. 
After all of the entries have been collected, they are saved into 
a CSV file, organized by tweet. Feature extraction and model 
training are based on these CSV files.

\subsection{HMM for Chinese Segmentation}\label{sect:chsegimpl}

As illustrated in Table~\ref{tab:chsegtable}, above, 
when segmenting a Chinese
sentence, we consider four states, namely, B for begin, M for middle, E for end, 
and~S for single. Thus, we train an HMM with four hidden states. 
The observation sequence consists of Chinese characters in the training dataset,
and the hidden states correspond to~B, M, E, and~S. 
It follows that the hidden state transition matrix of the
HMM is~$4\times4$ and of the form 
\begin{equation}\label{eq:HMMstateMatrix}
\begin{bmatrix}
\mbox{B}\to \mbox{B} & \mbox{B}\to \mbox{E} & \mbox{B}\to \mbox{M} & \mbox{B}\to \mbox{S}\\ 
\mbox{E}\to \mbox{B} & \mbox{E}\to \mbox{E} & \mbox{E}\to \mbox{M} & \mbox{E}\to \mbox{S}\\ 
\mbox{M}\to \mbox{B} & \mbox{M}\to \mbox{E} & \mbox{M}\to \mbox{M} & \mbox{M}\to \mbox{S}\\ 
\mbox{S}\to \mbox{B} & \mbox{S}\to \mbox{E} & \mbox{S}\to \mbox{M} & \mbox{S}\to \mbox{S}
\end{bmatrix}\mbox{.}
\end{equation}

We implemented this HMM-based Chinese text segmentation,
which is similar to that in~\cite{chseg1}. 
When training, the first character of each segmented word is marked 
as a beginning state (B). Then characters are marked as middle states (M),
until the last character is read, which is marked as an end state (E),
with any single-character words marked as such (S). 
The emission probability, 
the state transition probability, 
and the initial state probability 
are then used to update the state transition probability matrix 
in~\eref{eq:HMMstateMatrix}.
Subsequently, we use the trained HMM
to segment Weibo posts and comments 
line by line. 

\subsection{HMM for Emotion Classification}

For each word in a tweet or comment, we can calculate a three-dimensional 
vector based on its MI, CHI, and TF-IDF scores, 
as discussed in Section~\ref{sect:sentidata}, above. 
After calculating the feature vectors for each emotion, we obtain
a mean value of each feature over all tweets labeled by each specific emotion. 
This mean value is used as an observation. The transition feature
between states~$S_{k-1}$ and~$S_{k}$ is computed as
%%%%% ????? I still don't understand this, as it is either 0 and 1, not a probability
%%%%% Also, how does this involve emotions ?????
$$
P(S_{k} = s_{p}\,|\, S_{k-1} = s_{q}) = 
\left\{\hspace*{-0.035in}
  \begin{array}{cl}
    1 & \mbox{if } p=q+1\\ 
    0 & \mbox{otherwise}
  \end{array}
\right. ,
$$
where the HMM states~$S_k$ correspond to the features
MI, CHI, and TF-IDF. This
determines how close a feature vector in the test tweet is to 
those in the training set, with respect to the various emotions.
The emission probability
$$
  P(y_{k}\,|\,S_{k}^{e_{i}}) = J(y_{k}\,|\,S_{k}^{e_{i}}) = \frac{M_{11}}{M_{11}+M_{10}+M_{01}}
$$
can be calculated by Jaccard similarity~\cite{jaccard}, which 
measures the correlation between the feature vector~$y_{k}$ and 
the state~$S_{k}$, where~$M_{11}$ is the total number of tweets containing 
feature vector~$y_{k}$ and state~$S_{k}$ with respect to emotion~$e_{i}$,
$M_{10}$ is the number of tweets containing only state~$S_{k}$ 
with respect to emotion~$e_{i}$, and~$M_{01}$
is the number of tweets containing only feature vector $y_{k}$ with respect to emotion~$e_{i}$.

Table~\ref{tab:hmmemotiontable} gives an example of 
the relationship between three consecutive words~$w_0$, $w_1$, and~$w_2$
in a particular test case.
%of~$A$, $B$, and~$C$ in equation~\eref{eq:CHI}
%to the corresponding emotions.
%%%%%
%%%%% ????? Still don't understand what this means ?????
%%%%%
\begin{table}[!htb]
%    \caption{Features of terms~$A$, $B$, and~$C$ in equation~\eref{eq:CHI} for each emotion}
    \caption{Features for words~$w_0$, $w_1$, and~$w_2$ with respect to each emotion}
    \label{tab:hmmemotiontable}
    \vglue 0.1in
    \centering
%    \small
    \begin{tabular}{c|c|r|r|r} \midrule\midrule
        \textbf{Word} & \textbf{Emotion} & 
        		\multicolumn{1}{c|}{\textbf{MI}} & \multicolumn{1}{c|}{\textbf{CHI}} 
        		& \multicolumn{1}{c}{\textbf{TD-IDF}} \\ \midrule
%        \multirow{6}{*}{$A$} 
        \multirow{6}{*}{$w_0$} 
          & Happiness & 0.0012 & 0.0247 & 0.0009 \\
          & Anger & 0.0012 & 0.0247 & 0.0070 \\
          & Sadness & 0.0015 & 0.0100 & 0.0450 \\
          & Surprise & 0.0080 & $-0.0050$ & 0.0220 \\
          & Disgust & 0.0020 & 0.0470 & 0.0117 \\
          & Fear & 0.2200 & 0.0700 & 0.0009 \\\midrule
%        \multirow{6}{*}{$B$} 
        \multirow{6}{*}{$w_1$} 
          & Happiness & 0.0167 & 0.0064 & 0.1045 \\
          & Anger & $-0.0012$ & 0.0247 & 0.0009 \\
          & Sadness & 0.0200 & $-0.1416$ & 0.0009 \\
          & Surprise & 0.0012 & $-0.0247$ & $-0.0009$ \\
          & Disgust & $-0.0012$ & 0.0247 & 0.0009 \\
          & Fear & 0.0012 & 0.0247 & $-0.0009$ \\\midrule
%        \multirow{6}{*}{$C$} 
        \multirow{6}{*}{$w_2$} 
          & Happiness & 0.0012 & 0.0247 & 0.0009 \\
          & Anger & 0.0012 & 0.0247 & 0.0070 \\
          & Sadness & 0.0015 & 0.0100 & 0.0450 \\
          & Surprise & 0.3693 & 0.0820 & $-0.0119$ \\
          & Disgust & 0.0526 & 0.0247 & 0.0008 \\
          & Fear & 0.0012 & 0.0247 & 0.0007 \\
    \midrule\midrule
    \end{tabular}
\end{table}

We train an HMM for each of the six emotions. Then, we score a sample against
each model, and assign an emotion to the tweet based on the largest probability. 

%%%%% ????? This section seems confused ?????
%\subsection{Word2Vec Sentiment Score}
\subsection{Sentiment Score Calculation}

We first construct 
a Word2Vec model 
based on the~35,124 online shopping reviews 
contained in the sentiment analysis
dataset discussed in Section~\ref{sect:sentidata}. 
Note that this Word2Vec model
is based on segmented Chinese text. 
We use the GenSim 
package in Python~\cite{rehurek_lrec} to train this Word2Vec model.

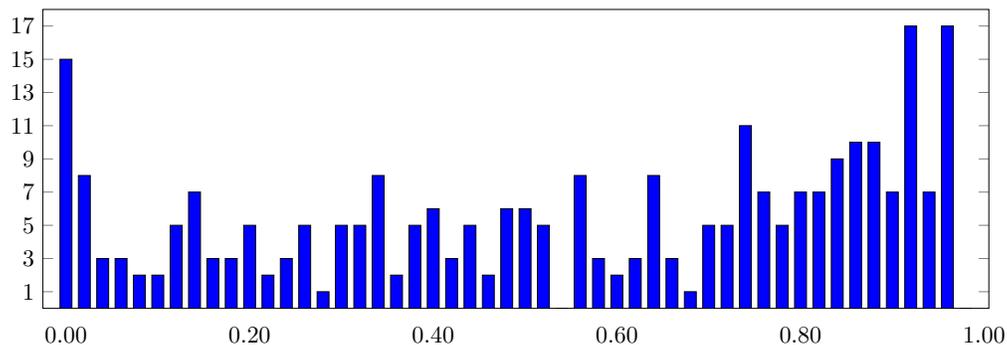
\begin{figure}[!htb]
    \centering
    \begin{tikzpicture}[scale=0.9, every node/.style={scale=1.0}]
    \begin{axis}[
        width  = 1.0*\textwidth,
        height = 6cm,
        xmin=-0.02,xmax=1.0,
        xtick={0,0.20,0.40,0.60,0.80,1.00},
        ymin=0.00,ymax=18.00,
        ytick={1,3,5,7,9,11,13,15,17},
        major x tick style = transparent,
        ybar=4*\pgflinewidth,
        bar width=5.0pt,
%        ymajorgrids = true,
%        xlabel = {Feature},
%        ylabel = {Weight},
%        symbolic x coords={5,
%        				      1,
%				      4,
%				      2,
%				      0,
%				      3,
%				      6,
%				      8,
%				      7},
	y tick label style={
    		/pgf/number format/.cd,
   		fixed,
   		fixed zerofill,
    		precision=0},
%	yticklabel pos=right,
%        xtick = data,
%        xtick = {0,1,2,3,4,5,6,7,8},
%	x tick label style={
%%		rotate=90,
%    		/pgf/number format/.cd,
%   		fixed,
%   		fixed zerofill,
%    		precision=0},
        x tick label style={
%        		rotate=60,
%		font=\footnotesize,
%		anchor=north east,
%		inner sep=0mm
    		/pgf/number format/.cd,
   		fixed,
   		fixed zerofill,
    		precision=2
		},
%		font=\small},
%        scaled y ticks = false,
	%%%%% numbers on bars and rotated
%        nodes near coords,
%        every node near coord/.append style={rotate=90, 
%        								  anchor=west, 
%								  font=\footnotesize,
%								  /pgf/number format/.cd,
%								  	fixed,
%									fixed zerofill,
%									precision=2
%								  },
        %%%%%
%        enlarge x limits=0.03,
        enlarge x limits=0.005,
        legend cell align=left,
        legend style={
%                at={(1,1.05)},
%                anchor=south east,
%	        nodes={rotate=90},%%%%% rotate text in legend
%                at={(0.125,0)},
%                at={(0.125,0)},
                at={(0.8775,0)},
                anchor=south,
                column sep=1ex
        }
    ]
\addplot[fill=blue,opacity=1.00]
coordinates {
(0,15)
(0.02,8)
(0.04,3)
(0.06,3)
(0.08,2)
(0.1,2)
(0.12,5)
(0.14,7)
(0.16,3)
(0.18,3)
(0.2,5)
(0.22,2)
(0.24,3)
(0.26,5)
(0.28,1)
(0.3,5)
(0.32,5)
(0.34,8)
(0.36,2)
(0.38,5)
(0.4,6)
(0.42,3)
(0.44,5)
(0.46,2)
(0.48,6)
(0.5,6)
(0.52,5)
(0.54,0)
(0.56,8)
(0.58,3)
(0.6,2)
(0.62,3)
(0.64,8)
(0.66,3)
(0.68,1)
(0.7,5)
(0.72,5)
(0.74,11)
(0.76,7)
(0.78,5)
(0.8,7)
(0.82,7)
(0.84,9)
(0.86,10)
(0.88,10)
(0.9,7)
(0.92,17)
(0.94,7)
(0.96,17)
(0.98,0)
};
%\legend{Level~1,Level~20}
\end{axis}
\end{tikzpicture}
    \caption{Sentiment score distribution for comments in Table~\ref{tab:alltrollfeature}, Row 1}
    \label{fig:jytsentiscore}
\end{figure}

Next, the resulting Word2Vec embeddings are used
to assign a sentiment score 
%%%%% This is synthetic data (in a sense). Seems a bit questionable
to segmented Chinese words, 
based on~3,286,543 tweets that we crawled from Weibo.
Thess sentiment scores are determined using \naive\ Bayes.
%For this sentiment score
%a~0 represents an extremely negative word, while~1 represents 
%an extremely positive word. 
%Note that this Word2Vec model is trained only on Weibo data.
%This is necessary in order to accurately reflect the casual 
%wording on Weibo, as compared to typical Chinese text.
Specifically, we use \naive\ Bayes to compute
$$
  P(c_{1}\,|\,w_{1},\ldots,w_{n}) = \frac{P(w_{1},\ldots,w_{n}\,|\,c_{1}) P(c_{1})}{
  	P(w_{1},\ldots,w_{n}\,|\,c_{1}) P(c_{1}) + P(w_{1},\ldots,w_{n}\,|\,c_{2}) P(c_{2})} ,
$$
where~$c_{1}$ represents the event that a word is positive
and~$c_{2}$ represents the event that a word is negative.
Our sentiment score is computed as the product of~$P(c_{1}\,|\,w_{1},\ldots,w_{n})$
and the word similarity score computed using the Word2Vec model.
%%%%% ?????????? What is the "word similarity score computed by the Word2Vec model" ?????
The resulting score can be viewed as the probability of a word
in a tweet being positive, where~0 represents an extremely negative word,
while~1 represents an extremely positive word. 
Figure~\ref{fig:jytsentiscore} illustrates 
a sample sentiment score distribution for one of our training datasets.
Note that this particular bar graph is based on the~812 comments 
corresponding to Weibo tweet ID~44275293, as listed in Table~\ref{tab:commentsdata}. 
Figure~\ref{fig:jytsentiscore} shows the sentiment score frequency 
count distribution for all~812 comments with brackets of width~0.02 
over the range of~0 to~1. 

%\begin{figure}[!htb]
%    \centering
%    \includegraphics[width=\textwidth]{images/jytsentiscore.png}
%    \caption{Sentiment score distribution for comments in Table~\ref{tab:alltrollfeature}, Row 1}
%    \label{fig:jytsentiscore}
%\end{figure}

\subsection{Troll Detection with XGBoost and SVM}\label{sect:XGB_SVM}

Our troll detection model is based on XGBoost.
We train our XGBoost models
using Python under the Jupyter Notebook environment.
For these XGBoosting troll detection models, we drop 
non-quantitative features, which leaves us with the features listed 
in Table~\ref{tab:originaltrolldetectionfeature}. Training an XGBoost model 
on all of these features, we achieve about~80\%\ accuracy.

\begin{table}[!htb]
    \caption{Troll detection statistics crawled from Weibo}\label{tab:originaltrolldetectionfeature}
    \vglue 0.1in
    \centering
%    \resizebox{\textwidth}{!}{
    \begin{tabular}{r|c|c} \midrule\midrule
           & \textbf{Feature} & \textbf{Description} \\ \midrule
        F0 & \texttt{follower} & Follower count  \\
        F1 & \texttt{following} & Following count  \\
        F2 & \texttt{original\_post} & Number of original tweets  \\
        F3 & \texttt{ffRatio} & \texttt{following} divided by  \texttt{follower} \\
        F4 & \texttt{foRatio} & \texttt{original\_post} divided by \texttt{follower}  \\
        F5 & \texttt{urank} & User activity rank in Weibo  \\
        F6 & \texttt{verified} & User certified or not  \\
        F7 & \texttt{description} & User's self description (1 or 0)  \\
        F8 & \texttt{freqComment} & User comments frequently or not  \\
        F9 & \texttt{like\_count} & Like count for comment  \\
        F10 & \texttt{floor\_number} & Location of comment \\
        F11 & \texttt{sentiment} & Sentiment score of the comment (0 to 1) \\
        F12 & \texttt{diffOriginalSenti} & \texttt{sentiment} minus sentiment of original \\
    \midrule\midrule
    \end{tabular}
%    }
\end{table}

\begin{figure}[!htb]
\centering
    \begin{tikzpicture}[scale=0.9, every node/.style={scale=1.0},rotate=-90]
    \begin{axis}[
        width  = 0.8*\textwidth,
        height = 10cm,
        ymin=0,ymax=130,
        ytick={0,20,40,60,80,100,120},
        major x tick style = transparent,
        ybar=4*\pgflinewidth,
        bar width=16.0pt,
%        ymajorgrids = true,
        xlabel = {Feature},
        xlabel style={rotate=180},
        ylabel = {Weight},
        yticklabel pos=right,
        symbolic x coords={F10,
        				      F11,
				      F2,
				      F1,
				      F9,
				      F6,
				      F0,
				      F3,
				      F5,
				      F7,
				      F4,
				      F8,
				      F12},
	y tick label style={
		rotate=90,
    		/pgf/number format/.cd,
   		fixed,
   		fixed zerofill,
    		precision=0},
%	yticklabel pos=right,
        xtick = data,
%        xtick = {0,1,2,3,4,5,6,7,8},
%	x tick label style={
%%		rotate=90,
%    		/pgf/number format/.cd,
%   		fixed,
%   		fixed zerofill,
%    		precision=0},
        x tick label style={
        		rotate=90,
%		font=\footnotesize,
%		anchor=north east,
%		inner sep=0mm
		},
%		font=\small},
%        scaled y ticks = false,
	%%%%% numbers on bars and rotated
        nodes near coords,
        every node near coord/.append style={rotate=90, 
        								  anchor=west, 
								  font=\footnotesize,
								  /pgf/number format/.cd,
								  	fixed,
									fixed zerofill,
									precision=0
								  },
        %%%%%
%        enlarge x limits=0.03,
        enlarge x limits=0.06,
        legend cell align=left,
        legend style={
%                at={(1,1.05)},
%                anchor=south east,
%	        nodes={rotate=90},%%%%% rotate text in legend
%                at={(0.125,0)},
%                at={(0.125,0)},
                at={(0.8775,0)},
                anchor=south,
                column sep=1ex
        }
    ]
\addplot[fill=blue,opacity=1.00]
coordinates {
(F10,115)
(F11,93)
(F2,68)
(F1,61)
(F9,55)
(F6,53)
(F0,28)
(F3,25)
(F5,25)
(F7,10)
(F4,8)
(F8,2)
(F12,0)
};
%\legend{Level~1,Level~20}
\end{axis}
\end{tikzpicture}
\caption{Initial XGBoost features ranking}
 \label{fig:featurerank1}
\end{figure}
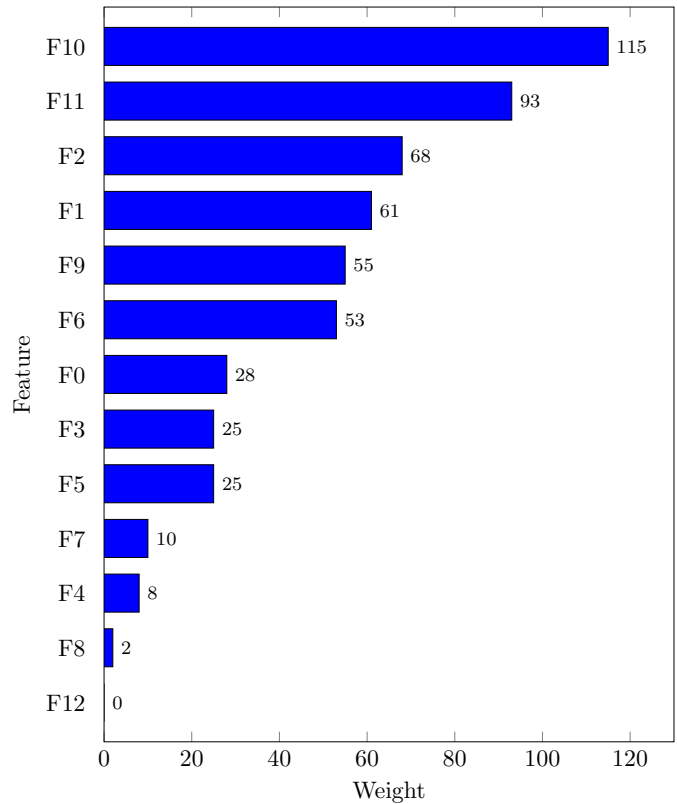

Ranking the features in this full-feature XGBoost model, we obtain the results in
Figure~\ref{fig:featurerank1}. Note that~F12 in Figure~\ref{fig:featurerank1} 
has a weight of~0, which implies that~F12 (the \texttt{diffOriginalSenti} feature)
contributed nothing to the classification.
%Based on these rankings, we eliminate features~F8 and~F12. 

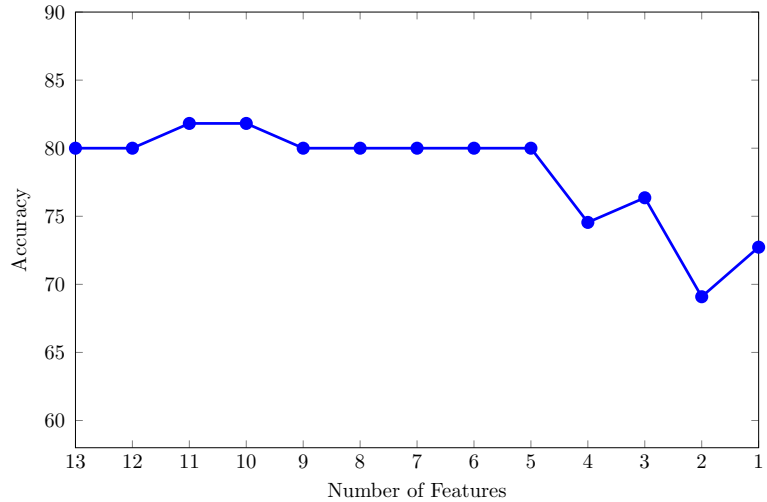
\begin{figure}[!htb]
\centering
    \begin{tikzpicture}[scale=0.65, every node/.style={scale=1.2}]
\begin{axis}[width=1.00\textwidth,
		   height=0.675\textwidth,
%		   /pgf/number format/1000 sep={},
	 	   x tick label style={
    		 	/pgf/number format/.cd,
   			fixed,
   			fixed zerofill,
    			precision=0},
	 	   y tick label style={
			xshift=-1.5pt,
    		 	/pgf/number format/.cd,
   			fixed,
   			fixed zerofill,
    			precision=0},
                    xmin=1,xmax=13,
                    x dir=reverse,
                    ymin=58.0,ymax=90.0,
                    legend pos=south west,
%                    nodes near coords,
%                    every node near coord/.append style={rotate=90, anchor=west, font=\footnotesize},
%                    every node near coord/.append style={anchor=west, font=\footnotesize},
                    xlabel={Number of Features},
                    ylabel={Accuracy}] 
\addplot[color=blue,ultra thick,mark=*,mark size=3.0] coordinates {
(1,72.73)
(2,69.09)
(3,76.36)
(4,74.55)
(5,80.00)
(6,80.00)
(7,80.00)
(8,80.00)
(9,80.00)
(10,81.82)
(11,81.82)
(12,80.00)
(13,80.00)
};
%\legend{High,Average,Low}
\end{axis}
\end{tikzpicture}
\caption{XGBoost model accuracy vs numbers of features}
\label{fig:trolldetectionmodelinitaccuracy}
\end{figure}

Next, we consider recursive feature elimination (RFE), 
where we drop the lowest ranked feature, then retrain the model.
Our RFE results are given in Figure~\ref{fig:trolldetectionmodelinitaccuracy}. 
We observe that the model improves when we drop the two
lowest ranked features, namely, F12 and~F8, but beyond that,
the model will lose accuracy if we drop additional features.
Hence, our optimal XGBoost model uses all of the features in
Table~\ref{tab:originaltrolldetectionfeature},
except~F8 and~F12. With this model, 
we achieve an accuracy of about~82\%.

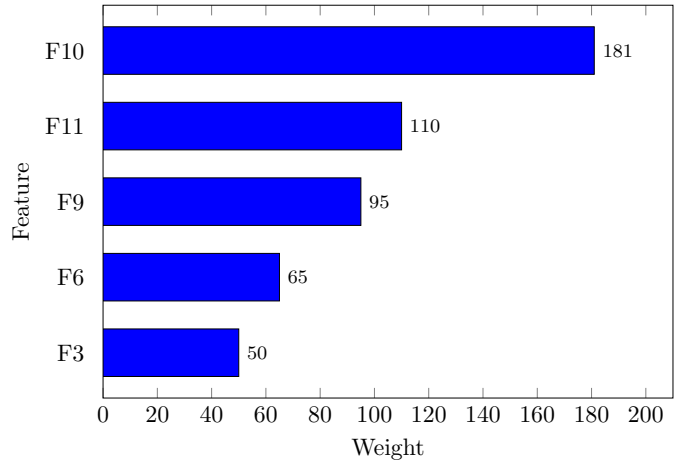
\begin{figure}[!htb]
\centering
    \begin{tikzpicture}[scale=0.9, every node/.style={scale=1.0},rotate=-90]
    \begin{axis}[
        width  = 0.475*\textwidth,
        height = 10cm,
        ymin=0,ymax=210,
        ytick={0,20,40,60,80,100,120,140,160,180,200},
        major x tick style = transparent,
        ybar=4*\pgflinewidth,
        bar width=20.0pt,
%        ymajorgrids = true,
        xlabel = {Feature},
        xlabel style={rotate=180},
        ylabel = {Weight},
        yticklabel pos=right,
%        symbolic x coords={F5,
%        				      F1,
%				      F4,
%				      F2,
%				      F0,
%				      F3,
%				      F6,
%				      F8,
%				      F7},
        symbolic x coords={F10,
        				      F11,
				      F9,
				      F6,
				      F3},
	y tick label style={
		rotate=90,
    		/pgf/number format/.cd,
   		fixed,
   		fixed zerofill,
    		precision=0},
%	yticklabel pos=right,
        xtick = data,
%        xtick = {0,1,2,3,4,5,6,7,8},
%	x tick label style={
%%		rotate=90,
%    		/pgf/number format/.cd,
%   		fixed,
%   		fixed zerofill,
%    		precision=0},
        x tick label style={
        		rotate=90,
%		font=\footnotesize,
%		anchor=north east,
%		inner sep=0mm
		},
%		font=\small},
%        scaled y ticks = false,
	%%%%% numbers on bars and rotated
        nodes near coords,
        every node near coord/.append style={rotate=90, 
        								  anchor=west, 
								  font=\footnotesize,
								  /pgf/number format/.cd,
								  	fixed,
									fixed zerofill,
									precision=0
								  },
        %%%%%
%        enlarge x limits=0.03,
        enlarge x limits=0.15,
        legend cell align=left,
        legend style={
%                at={(1,1.05)},
%                anchor=south east,
%	        nodes={rotate=90},%%%%% rotate text in legend
%                at={(0.125,0)},
%                at={(0.125,0)},
                at={(0.8775,0)},
                anchor=south,
                column sep=1ex
        }
    ]
\addplot[fill=blue,opacity=1.00]
coordinates {
%(F3,50)
%(F6,65)
%(F9,95)
%(F11,110)
%(F10,181)
(F10,181)
(F11,110)
(F9,95)
(F6,65)
(F3,50)
};
%\legend{Level~1,Level~20}
\end{axis}
\end{tikzpicture}
\caption{XGBoost features and rankings}\label{fig:featurerank2}
\end{figure}

The features used and their relative importance rank are given in 
Figure~\ref{fig:featurerank2}. Note that these are the features in
the XGBoost model.

For comparison with our XGBoost classifier, we also experiment with
an SVM classifier. We utilize the Python scikit-learn package~\cite{scikit} 
to train our SVM classifier. 
We compare the results of these SVM classification
experiments to the XGBoost experiments,
based over two different datasets and various sets
of features. Next, we summarize these results.

As discussed above, using XGBoost as our classification method
and RFE, we achieve an accuracy of about~82\%.
With some additional feature engineering,
we were able to increase this troll detection accuracy 
to~83.64\%\ on our Weibo crawled dataset,
using only the three features 
labeled as~F9, F10, F11 in Figure~\ref{fig:featurerank2}.
%see also Table~\ref{tab:originaltrolldetectionfeature}. 
%%%%% ????? The features mentioned here do not match those listed in Table 8 ?????
%These features correspond to \texttt{Following/Follower Ratio}, 
%\texttt{Original Post/Follower Ratio}, 
%and \texttt{Sentiment Score}. 
In addition, using our SVM classifier, we achieve~87.27\%\ accuracy 
on the same dataset, based on the same three features.
%using \texttt{Following/Follower Ratio}, \texttt{Original Post/Follower Ratio}, 
%and \texttt{Sentiment Score} as features. 
%The SVM model achieve mean AUC-ROC of 0.67409 in a 3 runs 5 splits ROC test. 

\begin{figure}[!htb]
\centering
    \begin{tikzpicture}[scale=0.825, every node/.style={scale=1.0}]
    \begin{axis}[
        width  = 0.6*\textwidth,
        height = 9cm,
        ymin=0,ymax=105,
        ytick={0,10,20,30,40,50,60,70,80,90,100},
        major x tick style = transparent,
        ybar=7.5*\pgflinewidth,
        bar width=30.0pt,
%        ymajorgrids = true,
%        xlabel = {Feature},
        ylabel = {Accuracy},
        symbolic x coords={XGBoost,
				      SVM
				      },
	y tick label style={
    		/pgf/number format/.cd,
   		fixed,
   		fixed zerofill,
    		precision=0},
%	yticklabel pos=right,
        xtick = data,
%        xtick = {0,1,2,3,4,5,6,7,8},
%	x tick label style={
%%		rotate=90,
%    		/pgf/number format/.cd,
%   		fixed,
%   		fixed zerofill,
%    		precision=0},
        x tick label style={
        		rotate=60,
		anchor=north east,
		inner sep=0mm
		},
%		font=\small},
%        scaled y ticks = false,
	%%%%% numbers on bars and rotated
        nodes near coords,
        every node near coord/.append style={rotate=90, 
        								  anchor=west, 
								  font=\footnotesize,
								  /pgf/number format/.cd,
								  	fixed,
									fixed zerofill,
									precision=2
								  },
        %%%%%
%        enlarge x limits=0.03,
        enlarge x limits=0.4,
        legend pos = south east,
        legend cell align=left,
        legend style={
%                at={(1,1.05)},
%                anchor=south east,
%	        nodes={rotate=90},%%%%% rotate text in legend
%                at={(0.125,0)},
%                at={(0.125,0)},
%                at={(0.8775,0)},
%                anchor=south,
                column sep=1ex
        }
    ]
\addplot[fill=blue,opacity=1.00]
coordinates {
(XGBoost,83.64)
(SVM,87.27)
};
\addplot[fill=red,opacity=1.00]
coordinates {
(XGBoost,89.00)
(SVM,81.82)
};
\legend{Weibo corpus,Review corpus}
\end{axis}
\end{tikzpicture}
\caption{Comparison of XGBoost and SVM}
\label{fig:accuracy}
\end{figure}
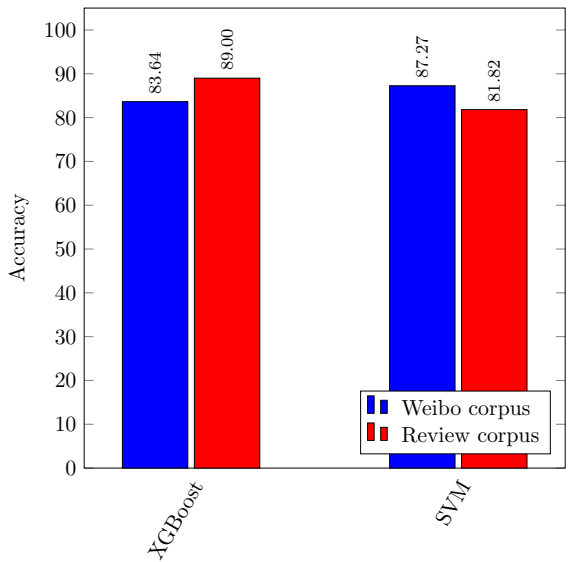

As another experiment, we compare our XGBoost and SVM
%%%%% What is this dataset????? Has this been mentioned before?????
models using the SnowNLP sentiment dataset for the sentiment score calculation. 
By using this training dataset for sentiment analysis, and with 
%%%%% ????? Again, these feature numbers do not match the names in Table 8
the addition of features~F3 
%(\texttt{User Rank}) 
and~F6,
%(\texttt{Floor Number}), 
the accuracy for XGBoost is~89\%. 
However, with this same dataset and feature set,
the SVM model achieves an accuracy of only~81.82\%.
These accuracy are summarized in the form of a bar graph in Figure~\ref{fig:accuracy}.

%\begin{figure}[!htb]
%\centering
%    \input figures/roc.tex
%\caption{Area under ROC curve for XGBoost}
%\label{fig:aucroc}
%\end{figure}

Our experimental results show that we can achieve an accuracy as high as~89\%,
which far exceeds the~78\%\ accuracy obtained in the
comparable previous work in~\cite{seah}, and matches the accuracy
in the previous work~\cite{weibospammer}.
A significant advantage of our approach is that it only requires a small number of 
easily obtained features, as compared to any previous research. 
This makes our troll detection technique highly efficient, and thus suitable for 
real-time troll detection, as validated by the Chrome extension
discussed in Section~\ref{sect:chrome}, above. 

\subsection{Chrome Extension for Troll Detection}\label{sect:chrome}

\begin{figure}[!htb]
    \centering
    \includegraphics[width=0.65\textwidth]{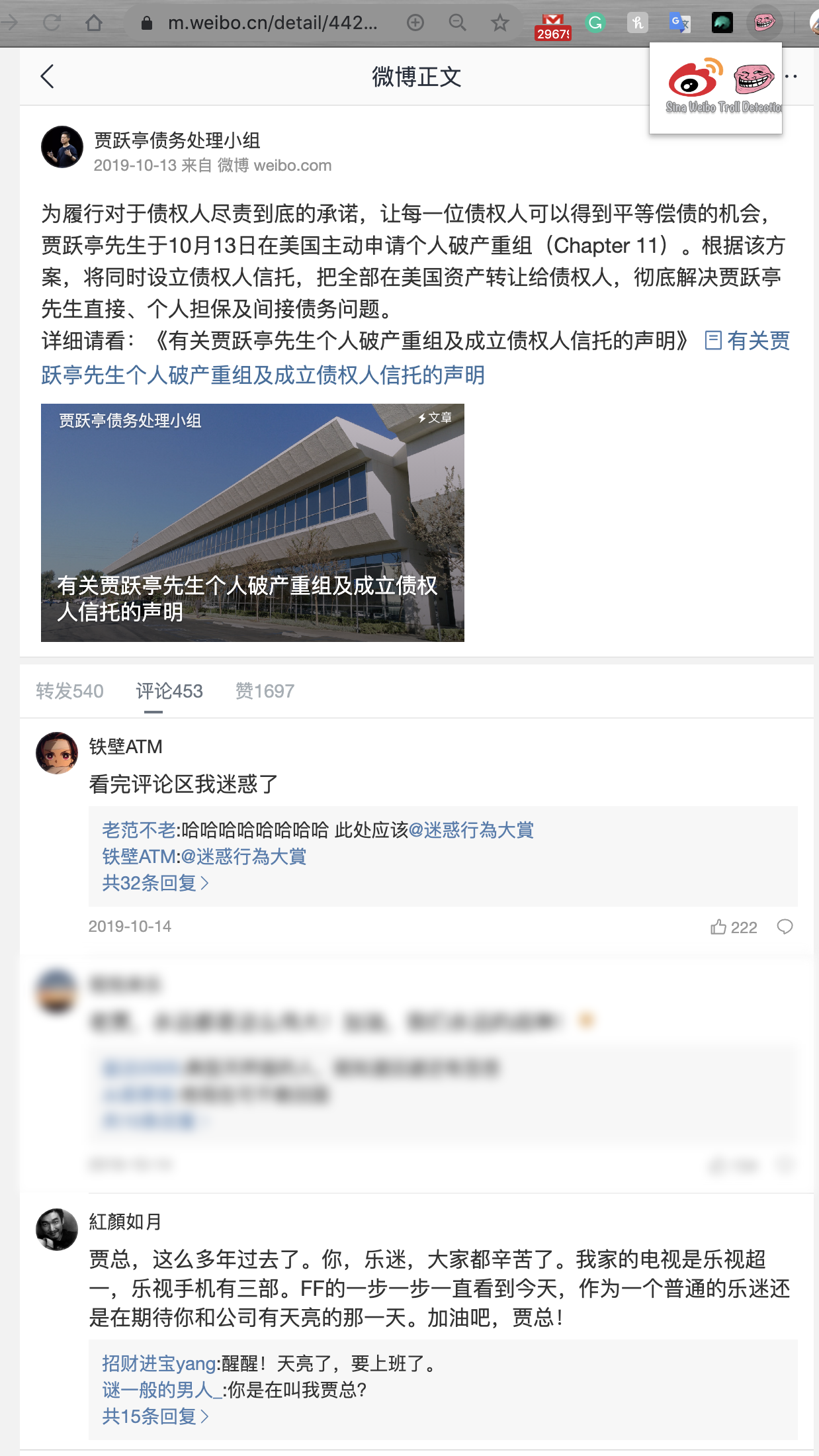}
    \caption{Chrome extension employing troll detection model}
    \label{fig:chromeext}
\end{figure}

Since our troll detection mechanism is written in Python, 
for real-time troll detection on the Sina Weibo mobile website,
we created a Chrome extension using HTML and JavaScript. 
In this extension,
we pass the JSON packet with Weibo comment information to the 
back-end, which is built on the Django framework~\cite{django}.
This back-end implements our troll detection model, as discussed above. 
The overall workflow for the plug-in is summarized as follows.
\begin{enumerate}
\item Run the crawler script against all the comments currently displayed 
in the browser under one tweet and send the packet to a server-side portal 
(currently running as localhost).
\item On the server-side, sort the essential user information and comment text 
from the returned JSON packet, as generated by the crawler.
\item On the server-side, run our sentiment analysis classifier 
against the comment text and acquire the text sentiment scores.
\item On the server-side, aggregate the sentiment scores and other user information,
feed these into our troll detection model, and return the troll detection result to the client-side plug-in.
\item Modify the CSS style sheet for any detected troll comments 
by adding an orange background behind the text.
\end{enumerate}
A screenshot showing this plug-in in action is given in Figure~\ref{fig:chromeext}.
Note that in this implementation, tweets and comments that have been flagged as 
potential troll activity are blurred.

\section{Conclusion and Future Work}\label{chap:conclusion}

The widespread use of social media enables information 
transfer to occur much faster than ever before. However, troll activities detract from
the utility of social media. Trolls have a variety of motivations, ranging from 
deception to profits, and it is not likely that these motivating factors
will diminish in the future. Therefore, intelligent defenses against trolls
are essential.

In this research, we utilized a variety of machine learning techniques to 
analyze comment content and user information on the Sina Weibo platform. 
By conducting sentiment analysis and by including user data aggregation, 
we were able to efficiently identify troll comments on Sina Weibo with 
higher accuracy, as compared to previous work.
We developed a Chrome extension that served to highlight the 
practicality of our approach.

For future work, more user data and other features can be considered.
In addition, deep learning techniques that utilize sequential information, such as 
long short term memory (LSTM) networks, could prove
useful. The Chrome extension that we have developed could be extended to 
support the Weibo desktop platform.

\bibliographystyle{plain}

%\bibliography{references.bib,Stamp-Mark.bib}
\bibliography{references.bib}

\end{document}